\documentclass{article}
\usepackage{authblk}       
\usepackage{amsmath, amssymb}
\usepackage{xfrac}           
\usepackage{fullpage}       
\usepackage{graphicx}      
\usepackage{placeins}        
\usepackage{parskip}          
\usepackage[T1]{fontenc}      
\usepackage[font=footnotesize]{caption}   
\usepackage{float}                  

\providecommand{\keywords}[1]
{
  \textbf{\textit{Keywords---}} #1
}

\begin{document}
\let\footnotesize\small
\date{}      
\title{Doomsday: Two Flaws}
\author[]{Mike Lampton}
\affil[]{Space Sciences Lab, UC Berkeley}
\affil[]{\textit{mlampton@berkeley.edu}}

\maketitle

\begin{abstract}  
\noindent
{\normalsize
Here I argue that the much-discussed Doomsday Argument (DA) has two flaws. Its mathematical 
flaw stems from applying frequentist probability or faulty Bayesian inference.  Its conceptual 
flaw is assuming that Copernican uniformity applies to a collection of ranked people and their ideas.  
I conclude that the DA has no predictive power whatsoever.}
\end{abstract}

\begin{center}
\keywords{Doomsday, Copernicanism, Bayes theorem}
\end{center}

\section{Introduction}   
The  ``Doomsday Argument'' (Carter 1983, Gott 1993, Leslie 1996)  applies a kind of Copernican 
principle that today's humankind does not occupy any special place in our long-term biological time span.  
With this hypothesis -- uniform random \emph{now}  -- there is only one parameter to be estimated, 
namely the total number $N$  of individuals  who will ever exist.  For any given size  $N$, this argument 
leads to a statistical confidence region for our future prospects, based entirely on the cumulative rank $R$ 
of individuals as of today: in only 1\% of trial universes under this hypothesis would inhabitants find 
$R<0.01 N$;  99\% of all such trial runs would have $R>0.01 N$.  Today our rank  $R \approx 10^{11}$, a
reasonable finding if $N=2 \times 10^{11}$ or even $10^{12}$.  So we should be  99\% confident that $N <10^{13}$, 
and doomsday is highly likely in the not-too-distant future.   

A concise statement of the DA was posed by John Leslie (1996) p.1:

\begin{quote}
    ``We ought to have some reluctance to believe that we are very exceptionally early, 
    for instance in the earliest 0.001 percent, among all humans who have ever lived.''
 \end{quote}

Its strengths and weaknesses have been discussed by many authors (Dieks 1992, 
Kopf et al 1994, Leslie 1996, Hanson 1998, Korb and Oliver 1998, Oliver and Korb 1998,
Bostrom 1999, Monton and Rousch 2001, Olum 2002, Sober 2003, Leslie 2008, 
Leslie 2010,  Gelman and Robert 2013,  Northcott 2016,  McCutcheon 2018); 
here I shall simply fix on what  I see as its two major errors.

\section{The Mathematical Flaws}  
To begin, where are we now?  Our present rank $R$ is about $10^{11}$ as I show in Figure 1 below. 
\begin{figure}[ht] 
\begin{center}
     \includegraphics{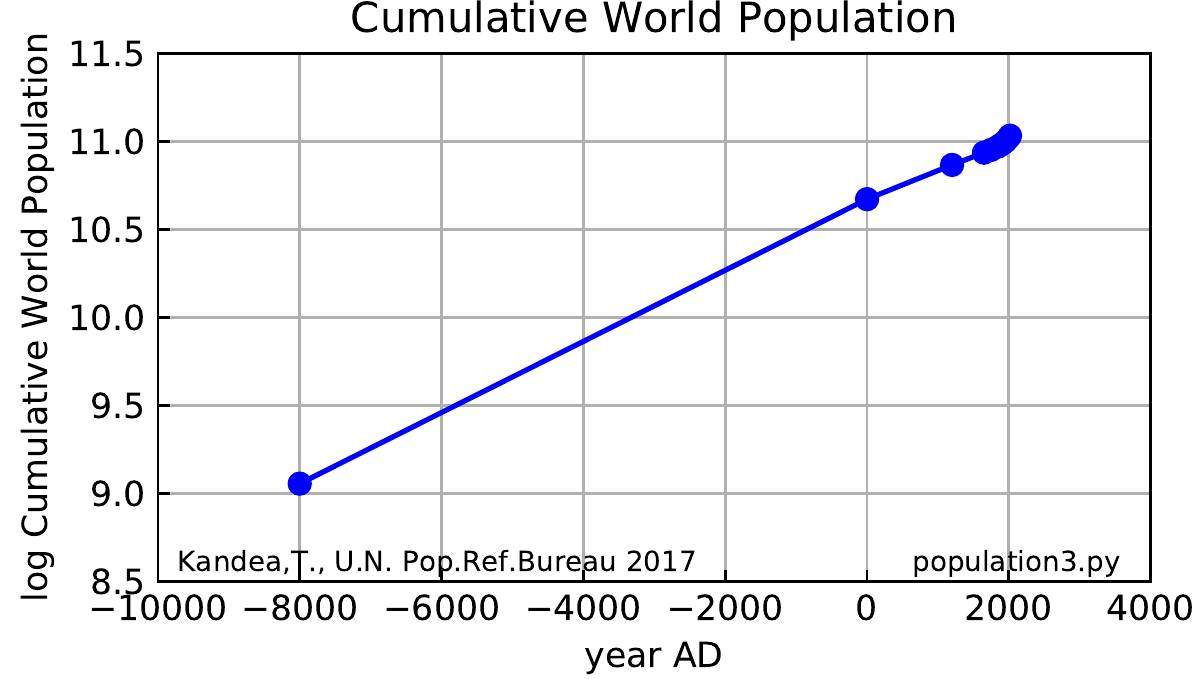}  
     \caption{Cumulative number of persons versus time, worldwide}
 \end{center}
\end{figure}
The earliest DAs applied a frequentist probability method to this rank.  Given $N$, there is only one chance in a 
million to have drawn a rank $R < 10^{-6} N$. True! But the frequentist DA then goes on to invert that, 
concluding that given an observed $R$, there is only one chance in a million that $N>10^6 R$,   
a circular argument. 

\pagebreak

Bayesian inference avoids this problem.
In Figure 2, I contrast frequentist vs. Bayesian inference by mapping how conditional probability depends on
population size $N$ and rank $R$ within that population.  
\begin{figure}[H]  
\begin{center}
     \includegraphics{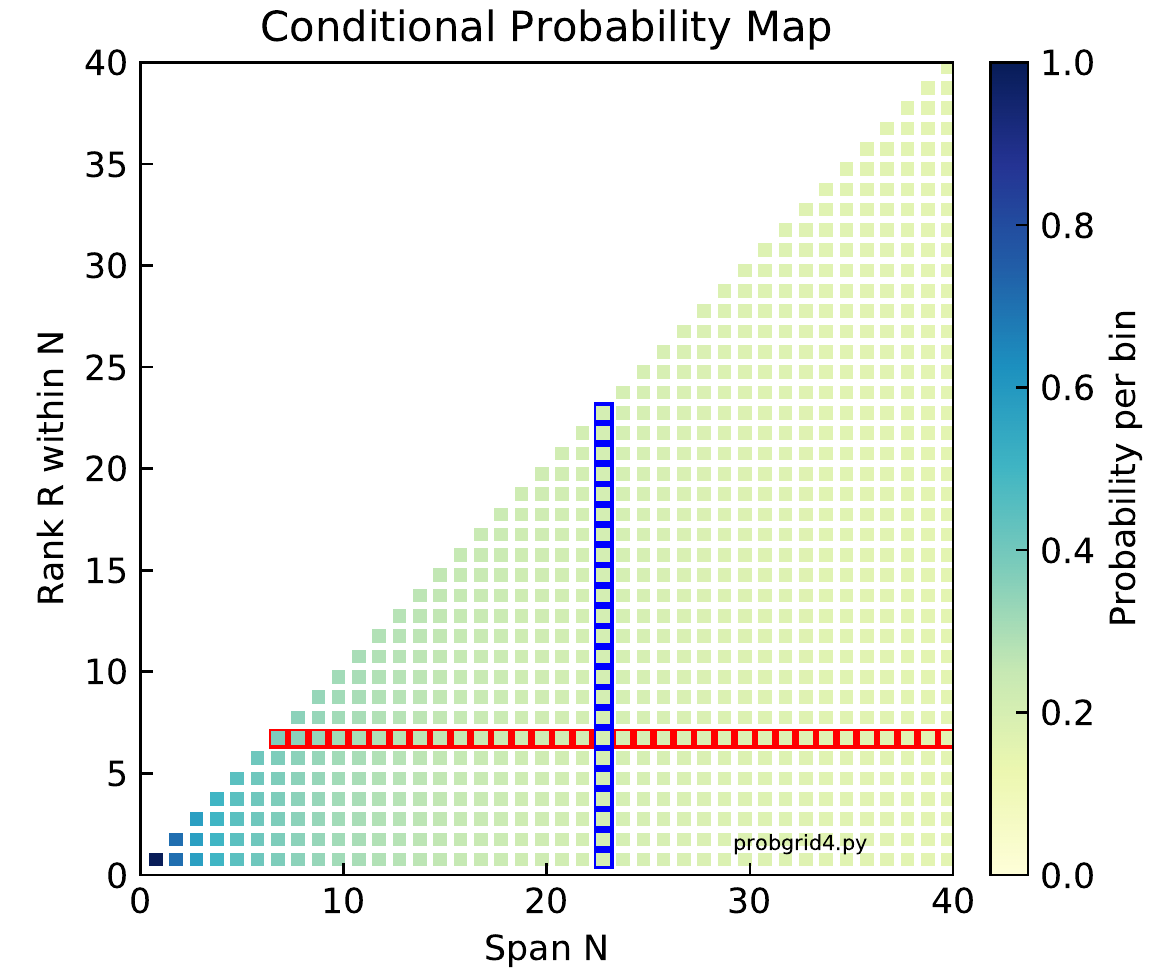}  
     \caption{Map of conditional probability $P(R \vert N)$.  }
\end{center}
\end{figure}
The frequentist view (vertical blue band) is a single column in this diagram:  given a cohort size $N$, the probability of
each ranked individual is $1/N$, equal for all ranks.  Being finite, it has simple confidence regions: you are 99\%
confident of not drawing a first percentile number.  The Bayesian view (red horizontal band) is a single row in this
diagram: we have drawn a 7, so what may we infer from this?  \emph{This is no longer a finite problem.} If all $N$ values 
have equal \emph{a priori} probabilities, we gather a likelihood of 1/7 from $N=7$, plus 1/8 from $N=8$, plus
1/9 from $N=9$, etc., all the way to infinity.    From this harmonic series we may now extract our composite hypotheses 
(for example, QD vs DD) for comparison.  Any decade above the observed rank contributes likelihood 2.303. 
The ratio of any pair of decades above the observed rank is 1.  
This is Rev. Bayes way of telling us that beyond our rank $R$, there is  no information here about the size of our 
cohort from a single sample. 

I will now apply Bayesian inference by decomposing the universe of possibilities using two numbers.  
Let each such civilization terminate with a finite cumulative number of members $N$, and for each 
civilization let each member be assigned a birth rank $R$ with $1<=R<=N$.  Next I will apply the 
clueless assumption: my own personal rank $R$ is uniformly distributed within the range $1<=R<=N$, 
and given $N$, my particular probability of having any one ticket number is $1/N$.
 
Finally I will adopt logarithmic scoring (McCutcheon 2019) and calculate the Bayes relative likelihood 
of two classes of futures:  QuickDoom ``QD''  $10^{11}<N<10^{12}$, versus DistantDoom ``DD'' with 
$10^{17}<N<10^{18}$.  Here I must distinguish two vastly different flavors of prior probability, although 
each is 1.00 when there is no prior information:
\begin{itemize}
    \item  $p(N)$ (small $p$) the prior probability of a simple hypothesis: a single $N$ value;
    \item  $P(G)$ (big $P$) the prior probability of composite hypothesis: a group $G$ of $N$ values.
\end{itemize}
 The groups QD and DD have a relative likelihood:
\begin{align}  
    \frac{L(QD)}{L(DD)} = \frac{P(QD) \cdot \sum_{10^{11}}^{10^{12}} \frac{1}{N} }{P(DD) \cdot \sum_{10^{17}}^{10^{18}} \frac{1}{N} }
       = \frac{P(QD)}{P(DD)}
\end{align}
For decade bins, each sum is 2.303 and they cancel.  The relative likelihood is just the ratio of our priors!  
Knowing only my rank, all I know about my cohort size is that it  exceeds my rank. Small, large: can't tell.  

The more recent DA formulations use Bayesian inference, as they should, but with flaws 
(Oliver and Korb 1998 examine many of these).
The simplest of the flawed Bayesian DA formulations is to compare one simple hypothesis, 
say QD=$10^{12}$ exactly, against  another simple hypothesis, say DD=$10^{18}$ exactly.  
No sums needed:
\begin{align}   
    \frac{L(QD)}{L(DD)} = \frac{p(QD) \cdot P(R \vert QD) }{p(DD) \cdot P(R \vert DD) }
       = 10^6 \cdot  \frac{p(QD)}{p(DD)}
\end{align}
For very reasonable equal priors, this tells us that QD is a million times more likely than DD.
Convincing, until you ask about the likelihood contributions from the $10^{12}$ nearby neighbors of QD
and the $10^{18}$ nearby neighbors of DD.  Shouldn't they count, as well?

One faulty way to include those neighbors is to average them over their respective realms.  To this I would
reply that averaging is inappropriate:  likelihoods add since each added simple hypothesis contributes
an additional way to obtain the observed rank.  

Another faulty way to include those neighbors is to assume that the prior of each future $N$ is known in 
advance from some previous measurement, and was (say) found to be $p(N)=1/N.$   This prior can then be 
installed into the composite summations, eliminating the group priors $P(X)$ in equation 1.  This gives:
\begin{align}   
    \frac{L(QD)}{L(DD)} = \frac{ \sum_{10^{11}}^{10^{12}} \frac{1}{N^2} }{\sum_{10^{17}}^{10^{18}} \frac{1}{N^2} }
       = \frac{1/10^{11}  \;  - \;   1/10^{12} }{1/10^{17} \;   - \;  1/10^{18}} =  10^6.
\end{align}
So QD is a million times more likely than DD, exactly what we put in with the priors. You can install
any preconceived priors, run the numbers, and get just what you installed.

My view is that \emph{this}  DA misunderstands priors.  The Bayesian process is to acknowledge that 
previous information exists about our two chosen alternative hypotheses.  From those priors
we use conditional probabilities to update our opinion in view of new data.  Since there is no prior 
information on what $N$ might be, we are obliged to regard all $p(N)$ values as equal, returning
us to equation 1.

\subsection*{The UrnFiller vs the UrnGuesser}
The UrnFiller (let's call her Jill) has two Urns (she calls them A and B) and into A she puts ten balls, 
numbered 1 through 10, and stirs them up.   Into B she puts 100 balls, numbered 1 to 100, and stirs them up.  
She is good at probability, and calculates that the chances of  some future guesser will draw a 7 is 
$P=0.5*0.1+0.5*0.01=0.055$ exactly: two possible choices between urns and then 1/10 or 1/100.  
She is right.  This is pure frequentist probability; no Bayesian inference needed.  

The UrnGuesser (let's call him Kent)  has a different problem.  He knows that Jill never lies, and 
is told the urns contain 10 and 100 balls. The urns of course look exactly the same.   He picks an urn.  
He is not allowed to weigh it.   He picks a ball.   It is 7.  Now what? There was a guaranteed 7 in each urn! 
So he has learned  nothing about his choice of urns.  Might have been the heavy urn, or the light urn.  
(Had he picked a 61, he would have nailed it,  but alas that didn't happen.)   

The UrnGuesser has learned about Bayesian inference, so applies it like this, to compare 
his relative likelihoods of two narrow categories Urn10 versus Urn100:
\begin{align}  
    \frac{L(10)}{L(100)} = \frac{p(10) * P(7 |10) }{p(100) * P(7 | 100)} = \frac{p(10) * 0.1}{p(100) * 0.01}
\end{align}
With this, he argues that it is ten times more likely that his chosen urn was the 10 than the 100. Those two sevens 
are ten times more dilute among 100 than they are under 10.  Being dilute means they are not going to be found 
so often.  A universe of ball-pickers should agree.  Between these two narrow categories, Bayesian inference gives us
perfect agreement with the UrnFiller's frequentist conclusion because either is the ratio of the two conditional
probabilities.  Problem Solved. QED.  End of discussion. 

But wait!  What if  \textit{what if there is only one urn}?  And what if \textit{the UrnFiller tells us nothing about N}?
The UrnGuesser picks a ball.   It is a 7.   How many balls were in the urn?  
This is a different problem, not a simple choice between two completely 
defined alternative narrow hypotheses, but between members of an infinite continuum of possible urns.  It is this switch that
forces us to abandon frequentist probability and adopt Bayesian likelihood methods. 

The general schema is to identify two groups of distinct, non-overlapping  hypotheses, add up the conditional 
probabilities of each simple hypothesis in each group, and compare their likelihoods.  Like this:
\begin{align}  
    \frac{L(A)}{L(B)} = \frac{\Sigma [p(A) \cdot P(7 |A)] }{\Sigma [p(B) \cdot P(7 | B)] }
\end{align}
To do this we must first decide what our groups will be. Having found a seven, it is certain that there were at 
least seven balls, but beyond that, the sky's the limit.  Let's create two alternative composite hypotheses: 
H70 spanning $7<N<70$ versus  H700 spanning $70<N<700$. 

For each $N$, before drawing any balls, the conditional probability of getting a 7 is simply $1/N$.  Since we 
have no reason to prefer one  $N$ value over another,  the \emph{a priori} factors are all 1, and vanish.
Our  integrated conditionals are just $ln(10) = 2.303$ in both cases.   So the likelihoods of the two 
cases are equal.  We have no reason to prefer one decade over the other.

Next, after drawing one ball, all decade spans above its rank $R$ remain equal. The \emph{a priori} 
factors are now zero for $N<R$, and remain 2.303 for decades above $R$, so the likelihood ratio is 1.0 for all test 
decades beyond $R$.    From one sample we learn  only that we may rule out $N<R$, which we knew already.    
Again, we have no reason to prefer one decade over the other, provided that both decades lie beyond $R$.   
We still have no upper bound on the size of our cohort.  In other words drawing the first sample
eliminates the smaller $N$ but does not affect the larger $N$ possibilities.

\subsection*{The Panzer Battle Commander vs the Allied Defense Commander} 
Our PBC knows he has fielded 1000 tanks and (unfortunately for him) each tank carries its serial number, 1...1000.   
If his tank \#77 were  captured in battle, that was in his first decile, but no surprise: it had to be in some decile, and all 
ten deciles are equally probable.  But if there were a second capture, say tank \#34, also in the first decile, that 
would be improbable: only a 10\%  that two captures would share a common bin. Extending this, the probability 
of $N$ captures from the first decile is $10^{-N}$, and the probability of $N$ captures all from a common 
decile is $10^{1-N}$.

Our ADC faces a very different problem.   He does not know $N$.   Had he captured a usefully large number 
$K$ of tanks and found a maximum serial number $M$, he would reasonably suppose that his captures 
span almost the entire production run and figure that $N$ is just a bit bigger than his $M$.
In the large-$K$ approximation, he would write $ \langle N \rangle \simeq M + AverageGap = M+M/K. $
However, this method fails for small $K$ (see Wikipedia 2019): the variance in  the $\langle N \rangle $  
estimator is infinite for $K<=3$;  indeed the mean $ \langle N \rangle$ is infinite if $K<=2$.   
One tank tells him squat about the size of its cohort. 

\subsection*{My Reply to John Leslie's ``Objection f''}  
In his thoughtful compendium of views on the DA, Leslie (1996, page 23) warns us:

\begin{quote}
 ``Don't object that there would be more chances of being born into a long-lasting human race, and that     
 these would precisely compensate for being born early in the history of that race.  The answer to this 
 is that there would be nothing automatically improbable in being in a short-lasting human race.  
 (Imagine that only ten people will ever have been born. Ought you to be specially surprised at finding 
 yourself among the ten?  No, for only those who are born can ever find themselves as anything.)''
\end{quote}

My view is that this objection conflates a fact with a distribution:

\emph{Draw one sample.}  Its observed rank is $R$.  Now $R$ has graduated
from being a distributed statistic to being a fact. We now have a firm lower limit on the population size 
$N>=R$, but still no upper limit.  For each remaining unobserved sample of rank $M$ we know $M \neq R$ and 
we can now write $P(M \vert N) = 1/(N-1)$:  $M$ remains uniformly distributed on $N$ except for the point $M=R$.
Yes, having drawn one sample, you still have a 50/50 chance
of being in the big urn cohort vs. the small urn cohort since one sample tells you nothing about the size of your
cohort.   But the $1/N$ or $1/(N-1)$ densities continue to apply to the remaining draws as yet unseen. 

My major point is that to adopt a single $N$, then obtain (say) a 99\% confidence region for $R$, then from our
observed $R$ to regard that $100 \times R$ as a 99\% confidence maximum on $N$, is a circular argument.  
It must be replaced with a confidence region obtained from Bayesian inference, and with just one sample that
region has no upper limit.

\section{The Uniformity Flaw}  
Figure 3 presents a linear pseudo-Copernican sequence of all humanity (discussed for example by Cirkovic and 
Balbi 2019).  (I term this pseudo-Copernican because I doubt that Copernicus himself would conflate the 
manifest growth in human thought with the spatial homogeneity of the universe.)
Suppose there are ten trillion  ($10^{13}$) individuals, ranked by birth.   Uniformly throughout, people 
formulate ideas randomly (spots).   Any idea can appear at any time, without any particular historical relationship 
to the many other ideas represented in the diagram.  Some are good ideas, some are bad. One of them (red) might
be the Doomsday Argument.
\begin{figure}[ht] 
 \begin{center}
     \includegraphics{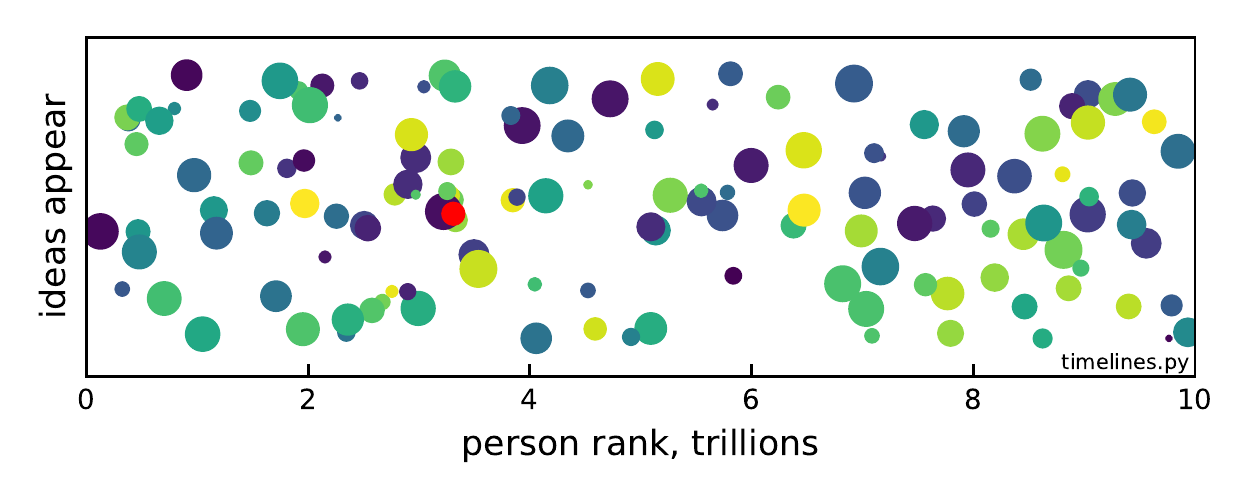}  
         \caption{A cartoon illustrating a statistically uniform distribution of people and their ideas. }
 \end{center}
\end{figure}

In contrast, I show in Figure 4 a growth of knowledge on a logarithmic scale.  I do not regard ideas as unrelated.  
They build on previous ideas; they become confirmed or modified or rejected.  
\begin{figure}[ht] 
 \begin{center}
     \includegraphics{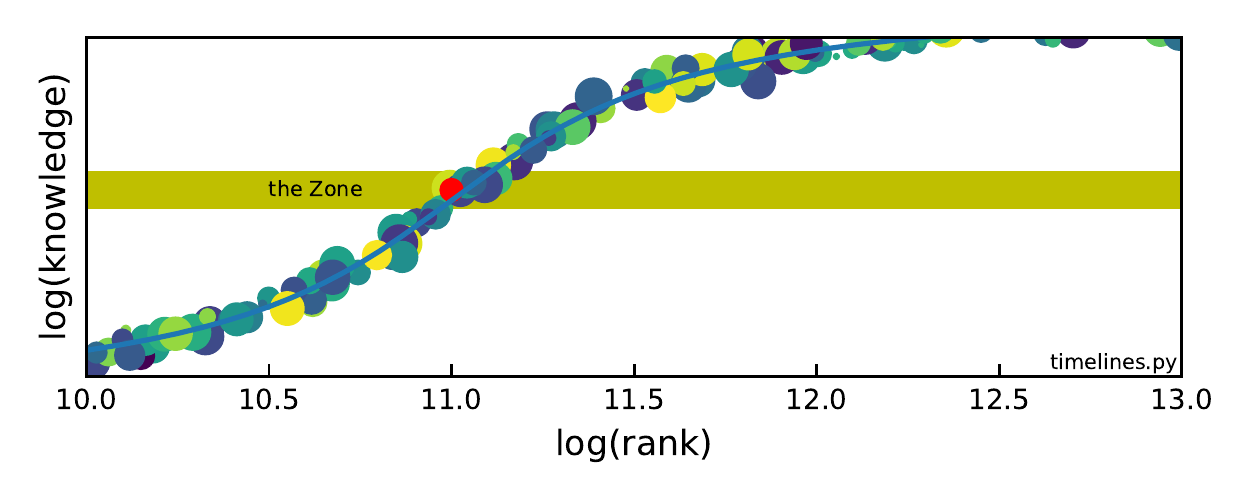}  
         \caption{Logarithmic scale cartoon showing the growth of knowledge as relationships between
         previous ideas. }
 \end{center}
\end{figure}
So, I plot knowledge as a curve of networked ideas enjoying a mostly steady rise. 
In particular, the discovery of the DA depends on 
writing, mathematics, thought experiments, and probability\footnote{\begin{small}This 
rise need not depend on any improvements in human intelligence. Gould and Eldredge (1977)  describe 
evolutionary trends as punctuated equilibria: long periods without significant genomic change, interrupted 
by relatively rapid shifts in their properties.  Quite likely, human intelligence has followed this \emph{punk eek} 
pattern. Bostrom (2009) compares long-term static intelligence versus various slow and rapid improvements.  
One extreme is the singularity (Ulam 1958; Good 1965; Vinge 1993; Sandberg 2009) that propels human 
capabilities far beyond those experienced today.  My take on these possibilities is that --- even without them --- 
growth in aggregate human knowledge will assure  a much improved likelihood of overcoming future crises.  
\end{small}}.   At our present rank we have recently
entered the Zone: we have invented the DA, we are discussing it, and we may soon outgrow it.

A dedicated Doomsdayist would immediately object that our Zone is at rank $R\simeq10^{11}$ because that is reasonably 
situated among $10^{11}..10^{12}$ civilizations, whereas long-lived Distant Doom civilizations, (say) those with $R \simeq 10^{17}$,  
would statistically be unlikely to find their Zones around $10^{11}$: they are far more likely to be encountered at 
random locations throughout their much larger spans.  

To this, I have two responses.  First, the Zone is systematic, not random, and its arrival rank is set by the historical development of 
essentials like writing, teaching, and mathematics, all of which come from the past. They are utterly independent of 
future developments. Statistical methods do not apply.  Second, it is an error to apply conditional probability 
$P(R \vert N)$ to a question demanding inferential likelihood $L(N \vert R)$.

\section{Conclusion}  
Two extreme cases -- a modestly future-human and an outrageously future- or post-human 
scenarios --- cannot be mathematically distinguished by knowing only our present population size. 
Humanity's passenger train definitely starts out finite in size, but might outgrow 
any such limit as time progresses.

\emph{Acknowledgment:}  I gratefully acknowledge having illuminating discussions with 
John Leslie and Ken Olum on the many facets of the DA.  My errors are of course mine alone.

\section*{References}

Bostrom, N., "The Doomsday Argument is Alive and Kicking," Mind v.108, pp.539-550 (1999).

Bostrom, N., "Future of Humanity," New Waves in Philosophy of Technology, eds. 
Jan-Kyrre Berg Olsen, Evan Selinger, and Soren Riis (New York: Palgrave McMillan, 2009).

Carter, B. "The anthropic principle and its implications for biological evolution," Phil. Trans. Royal Soc 
London A310, pp.347-363 (1983). 

Cirkovic, M. M., and Balbi, A., "Copernicanism and the typicality in time,"  IJA (2019).

Dieks, D., "Doomsday --- or: the Dangers of Statistics,", Philosophical Quarterly v.42, Issue 166, pp.78-84 (1992).

Gelman A. and Robert C.P., "The Perceived Absurdity of Bayesian Inference," Amer. Stat. Assoc. v.67 No.1, pp.1-5 (2013). 

Good, I.J., "Speculations Concerning the First Ultraintelligent Machine", Advances in Computers, vol. 6 (1965).
 
Gott, J.R., "Implications of the Copernican Principle for our future prospects," Nature v.363, 27 (1993). 

Gould, S.J., and Eldredge, N., "Punctuated equilibria," Paleobiology v3 No.2, pp.115-151, (1977). 

Hanson, R., "Critiquing the Doomsday Argument," (1998), https://web.archive.org

Kopf, T., Krtous, P., and Page, D., "Too Soon for Doom Gloom?" arXiv 9407002 (1994).

Korb, K.B., and Oliver, J., "A Refutation of the Doomsday Argument," Mind v.107 pp.403-410 (1998).

Leslie, J., "The End of the World," London and New York: Routledge, (1996).  

Leslie, J.., "Infinitely Long Afterlives and the Doomsday Argument?," Philosophy Vol.83 No.326 pp.519-524 (2008).

Leslie, J., "Risk that Humans will soon be extinct,"  Philosophy v85 No.4 (2010). 

McCutcheon, R., "What, Precisely, is Carter's Doomsday Argument?" https://philpapers.org, (2018)

McCutcheon, R., "In Favor of Logarithmic Scoring," Phil. Sci. v86 No.32  286-303 (2019)

Monton B. and Roush S., "Gott's Doomsday Argument,"  https://philpapers.org (2001).

Northcott, R., "A Dilemma for the Doomsday Argument," https://philpapers.org (2016).

Oliver, J. J., and Korb,  "A Bayesian Analysis of the Doomsday Argument,"  Monash U. (1998).

Olum, K., "The Doomsday Argument and the Number of Possible Observers," arXiv gr-qc-0009081 (2001) 
and Phil.Quarterly v.52 No.207 pp.164-184 (2002).

Sandberg, A., "An Overview of Models of the Technological Singularity," agi-conf.org, (2009).

Sober, E., "An Empirical Critique of Two Versions Of The Doomsday Argument," Synthese v.135 (2003).

Ulam, S., "Tribute to John von Neumann,"" Bull.Amer.Math.Soc. v64 No.3 pp1-49 (1958)

Vinge, V., "The Coming Technological Singularity: How to Survive in the Post-Human Era,"
NASA Publication CP-10129 (1993).

Wikipedia,  https://en.wikipedia.org/wiki/German-tank-problem (2019)

\end{document}